\documentclass[preprint2]{aastex}

\usepackage[utf8x]{inputenc}
\usepackage[T1]{fontenc}
\usepackage{ae}
\usepackage{float}
\usepackage[multidot]{grffile}
\usepackage{xcolor}
\usepackage{graphicx}
\usepackage{sistyle}
\usepackage{amsmath}
\usepackage{amsfonts}
\usepackage{amssymb}
\usepackage{arydshln}
\usepackage{dsfont}
\usepackage[bottom,hang]{footmisc}
\usepackage{booktabs}
\usepackage{multicol}

\DeclareMathOperator{\ZETA}{\zeta}

\newcommand{\bs}{\begin{split}}
\newcommand{\es}{\end{split}}
\newcommand{\be}{\begin{equation}}
\newcommand{\ee}{\end{equation}}
\renewcommand{\ni}{\noindent}
\newcommand{\dd}{\mathrm{d}}
\renewcommand{\d}{\mathrm{d}}

\newcommand{\e}{\,\mathrm{e}}
\renewcommand{\in}{\mathrm{i}}
\renewcommand{\i}{\,\mathrm{i}\,}

\newcommand{\gvec}[1]{\boldsymbol{#1}}
\renewcommand{\vec}[1]{\mathbf{#1}}
\newcommand{\bea}{\begin{eqnarray}}
\newcommand{\eea}{\end{eqnarray}}
\newcommand{\ew}[1]{\left \langle #1 \right \rangle}
\newcommand{\abs}[1]{\left \lvert\,  #1 \,\right \rvert }
\newcommand{\brakket}[3]{\left \langle #1 \left |\,  #2 \, \right | #3 \right \rangle}
\renewcommand{\epsilon}{\varepsilon}
\newcommand{\f}{\mathrm{f}}
\newcommand{\fin}{\mathrm{fi}}

\newcommand{\ket}[1]{ \left | #1 \right \rangle }

\newcommand{\braket}[2]{\left \langle #1 \right | \left. \! #2 \right \rangle}
\renewcommand{\epsilon}{\varepsilon}
\renewcommand{\vec}[1]{\mathbf{#1}}
\renewcommand{\d}{\text{d}}

\begin{document}

\title{Hartree-Fock calculations for the photoionisation of helium and helium-like ions in neutron star magnetic fields}

\author{Thorsten Kersting, Damir Zajec, Peter Diemand, G\"unter Wunner} 
\affil{Institut f\"ur Theoretische Physik 1, Universit\"at Stuttgart, 70550 Stuttgart, Germany}
\email{kersting@itp1.uni-stuttgart.de}


\begin{abstract}
We derive the photoionisation cross section in dipole approximation for many-electron atoms and ions for neutron star magnetic field strengths in the range of $10^7$ to $10^9$ T.
Both bound and continuum states are treated in adiabatic approximation in a self-consistent way. Bound states are calculated by solving the Hartree-Fock-Roothaan equations using finite-element and $B$-spline techniques while the continuum orbital is calculated by direct integration of the Hartree-Fock equations in the mean-field potential of the remaining bound orbitals.
We take into account mass and photon density in the neutron star's atmosphere, finite nuclear mass as well as thermal occupation of the levels.
The data may be of importance for the quantitative interpretation of observed x-ray spectra that originate from the thermal emission of isolated neutron stars.
They can serve as input for modeling neutron star atmospheres as regards chemical composition, magnetic field strength, temperature, and redshift.
Our main focus in this paper lies on helium and helium-like oxygen. These two-electron systems are simple enough to calculate all possible transitions when limiting the quantum numbers and should show all the basic structures and behaviour of other two-electron systems up to iron. 
\end{abstract}

\maketitle

\section{Introduction}

Atoms in neutron star magnetic fields have been the subject of research for almost 40 years, starting with the analysis of the simplest chemical element, the 
hydrogen atom.
A historical review of the work on the hydrogen atom in strong magnetic fields can be found in the book by Ruder et al. \citep{Ruder}. In particular, in the early 1980's bound-bound \citep{Ruder} as well as bound-free \citep{Schmittetal} transitions of hydrogen in strong magnetic fields were calculated. 

Studies of heavier elements have received renewed impetus from the discovery of 
broad absorption features in the thermal X-ray emission spectra of the isolated
neutron star 1E 1207.4-5209 \citep{sanwal2002,mereghetti2002,bignami04a} and three other isolated neutron stars \citep{haberl2003, haberl2004, kerkwijk2004} by the Chandra X-Ray Observatory of NASA and the XMM-Newton Observatory of ESA. These
features could be of atomic origin  \citep{Mori2002,mori2005,Mori2006a}.  
As can be seen from the review article  on neutron
star thermal emission by Zavlin \citep{Zavlin09a}, magnetized models of neutron star 
atmospheres  have largely been confined to hydrogen atmospheres, partially or
fully ionised. 
There were just a few attempts in the literature to model
magnetized heavy element (carbon, oxygen, neon, iron) atmospheres. One attempt
was undertaken by Mori and Ho \citep{Mori07a}. They showed that the features
observed in 1E\;1207 could be due to transitions of oxygen in different
ionisation stages. Pavlov and Bezchastnov \citep{Pavlov05a} 
suggested that the features
can be produced by bound-bound transitions of singly ionised helium
in strong magnetic fields. 
As noted by Zavlin and Pavlov \citep{Zavlin2002}, Rajagopal et al \citep{Rajagopal97} tried to model an iron atmosphere "`with the use of rather crude approximations for the very complicated properties of iron ions in strong magnetic fields."'
A further interpretation in terms of peaks in the
energy dependence of the free-free opacity has been put forward
by Suleimanov et al. \citep{Suleimanov10a}, but again their modelling 
is restricted to hydrogen atmospheres.

In view of the unknown chemical compositions of the atmospheres of neutron
stars, any possible fusion product, i.e., all elements from hydrogen to iron in various ionisation stages, could produce features in the thermal emission spectra of neutron stars. In order to be in a position to calculate synthetic spectra
for models of neutron star atmospheres and to compare with observed spectra,
in principle a complete knowledge of both the bound-bound transitions and the bound-free transitions
of all these elements in strong magnetic fields, and in all ionisation
stages, would be required.
While bound-bound transitions in  many-electron systems in neutron star
magnetic field strengths 
have already been investigated in the literature \citep{Engeletal}, calculations
of photoionisation cross sections are few. Medin, Lai, and Potekhin \citep{Medin}
calculated a small number of bound-free transitions for helium in strong magnetic fields, as did Mori and Hailey \citep{Mori2006a} for oxygen and neon ions.

We wish to extend these calculations by performing a systematic and
complete study of 
the total photoionisation cross sections of two-electron atoms for
the physically realistic situation that the initial 
bound states are thermally occupied, and the atoms are exposed to black body
radiation of the same temperature. As in \citep{Medin} we work with the
adiabatic approximation 
\citep{SchiffSnyder,CanutoKelly} which implies that all electrons
are in the lowest Landau level ($n=0$), with different magnetic
quantum numbers ($m = 0, -1, -2, -3, \cdots$), and the parts of the wave 
functions along the magnetic field are calculated self-consistently.
(Adiabatic means, intuitively speaking, that the motion of the electrons
in the plane perpendicular to the field is fast compared to the motion along the
field, in which the electrons only feel the Coulomb potential of the nucleus,
averaged over Landau orbitals.) 
In the low filed regime it is absolutely necessary to perform non-adiabatic calculations for bound states, as has been done, e.g. by \citep{Becken99,Schi2012} but in the range of magnetic fields considered in this paper, the adiabatic approximation is applicable for bound states. Also the unbound state is accurately described in adiabatic approximation despite the strength of the magnetic field as long as its energy does not exceed the landau excitation threshold. The only effect appearing in a non-adiabatic treatment, is resonance of higher Landau-levels for bound-free transitions \citep{Potekhin97}. But these resonances are much too high to be thermally excited ($\approx 10$\,keV). 

The specific difficulty in calculating photoionisation cross sections 
of many-electron atoms in strong magnetic fields lies in the lacking shell structure of these atoms. For every magnetic quantum number $m$ possible in the
lowest Landau level there exists an energetically strongly lowered ground 
state (tightly bound state), while the energies of the excited states
for every $m$ form a Rydberg series (hydrogen-like states). Thus all
excited states are practically $m$-degenerate, and, assuming thermal
occupation of the initial states, occupied with equal probability. 
The summation of the individual cross sections for the
ionisation of the atoms from the $m$-degenerate states by photons of 
given energy  would therefore diverge. 
We will introduce a physically reasonable 
cut-off criterion to avoid the divergence. 

It is the purpose of this paper to demonstrate that we have developed a powerful tool for calculating practically complete photoionisation cross sections without the need to compromise accuracy for speed. These cross sections can serve as input for the computation of opacities and for more detailed astrophysical modelings. 

The organisation of the paper is as follows:
In Sec. \ref{atoms}, we briefly recapitulate the properties of atoms in intense magnetic fields and review the Hartree-Fock-Finite-Elements Roothaan (HFFER) method used to determine the bound initial states. 
The orbitals of the resulting ion and of the continuum electron after the photoionisation are calculated separately and then combined to a Slater determinant characterising the final state.
The cross sections are derived in Sec. \ref{cross}.
Atomic data for helium and helium-like oxygen are presented in Sec. \ref{results}. The influences of photon and mass density, finite nuclear mass and thermal occupation are discussed in Sec. \ref{influences}.

\section{Atoms in very strong magnetic fields}\label{atoms}

For the readers' convenience we briefly review the basics of atoms in strong 
magnetic fields \citep{Ruder}. A magnetic field $B$ is called strong if it 
exceeds the  characteristic field strength $B_0=4.70108\times10^5$\,T, 
i.e., when the magnetic field parameter  $\beta = B/B_0$ is large. In our
calculations we will concentrate on a magnetic field strength of $B = 10^8$\,T
which corresponds to $\beta \approx 212$. At $B_0$ the Larmor radius $a_{\rm L}$
becomes equal to the Bohr radius $a_{0}$. We note that $a_{\rm L} = a_{0}/\sqrt{\beta}$.

In strong fields the cylindrical symmetry of the magnetic field dominates,
and it is therefore appropriate to work in cylindrical coordinates (we
assume the field to point in the $z$ direction). Then we can write the 
Hamiltonian for
an $N$-electron atom or ion with nuclear charge $Z$ (in atomic Rydberg units, i.e., $\hbar = 4\pi \epsilon_0 = 2 m_e = e^2/2 \equiv 1$) 
in the form 
$
H = \sum_i^N \hat{H_{i}}\,
$, 
with

\begin{multline}
 \hat{H_{i}}=-\left(\frac{\partial^2}{\partial\rho_i^2}+\frac{1}{\rho_i}\frac{\partial}{\partial\rho_i}+\frac{1}{\rho_i^2}\frac{\partial^2}{\partial\phi_i^2}+\frac{\partial^2}{\partial z_i^2}\right)\mathds{1} 
 \\
 -2\i\beta\frac{\partial}{\partial\phi_i}\mathds{1} +\beta^2\rho_i^2\mathds{1}+2\beta\hat{\sigma}_{z,i}-\frac{2Z}{|\vec{r}_i|}\mathds{1} 
 \\
 + \frac{1}{2} \sum_{j\neq i}^{N} \frac{2}{|\vec{r}_{i}-\vec{r}_{j}|}\mathds{1}\, 
\label{hamiltonzylinder}
\end{multline}
being the Hamiltonian for the $i$-th electron. In (\ref{hamiltonzylinder})
$\hat{\sigma}_{z}$ is the $z$ Pauli-matrix, and the term describes the interaction of the
spin magnetic moment with the external field. At the magnetic field strength
we consider, $B= 10^8$\,T, the splin flip energy, which is also the 
Landau excitation energy, amounts to $11.6$~keV, which is
much larger than any single-orbital energy that we will encounter. Therefore
it is justified to assume all electron spins to be aligned antiparallel 
to the magnetic field, and all single-electron states to occupy the 
lowest Landau level $n=0$. The spin part of the state will be omitted in
what follows.

The single-particle orbital of electron $i$ is given by 
\begin{equation}\label{opo}
  \Psi_i(\rho,\phi,z)=\Phi_{m_i}(\rho,\phi)g_{m_i\nu_i}(z)\, ,
\end{equation}
\noindent
where $\Phi_{m_i}(\rho,\phi)$ is the lowest Landau orbital, with magnetic quantum
number  $m_i$, and $g_{m_i\nu_i}(z)$ denotes the longitudinal part of the wave function. 
The quantum number  $\nu_i$ counts the number of nodes of the wave function 
along the $z$ direction. In the following, states with quantum numbers $m$ and $\nu$ will be denoted by $(m, \nu)$. Recall that in the lowest Landau level $m$
only assumes non-positive values, $m=0, -1, -2, -3, \cdots$.

From the single-electron orbitals (\ref{opo}) of the bound electrons  
a Slater determinant is constructed, and the wave functions $g_{m_i\nu_i}(z)$ 
are determined self-consistently
by solving the resulting Hartree-Fock equations. In our calculations,
the $z$ axis is divided into finite elements, and the longitudinal wave 
functions are expanded in terms of $B$-splines. The $B$-spline
expansion coefficients are determined by minimising the total energy functional
(for computational details of this HFFER method cf. \citep{Eng2009,Schi2012}).

The final state of the bound-free transition consists of the ionised core and 
the electron which is photoionised into the continuum. 
Near the atomic core the unbound electron is influenced by the bound electrons,
while the bound electrons can be assumed to be unaffected by the unbound 
electron \citep{Medin}. Therefore the initial state of the atom 
and the bound part of the final state, i.e., the ionised core, can be 
calculated separately in advance by the above mentioned method.

The wave function of the continuum electron is calculated by solving
the Hartree-Fock equations that result from its interaction with the
electrons of the remaining core:

\begin{multline}
\biggl[-\frac{\partial^{2}}{\partial z^{2}}+V_{m}(z)  - E' 
\\
  + \sum_{j=1}^{N-1}\int \!\! \dd z'|g_{m_{j}\nu_{j}}(z')|^{2}U_{m_{j}m}(z,z') \biggr] g_{mE'}(z) 
\\
=\!\! \sum_{j=1}^{N-1}   \int \!\! \dd z' g_{mE'}(z')g_{m_{j}\nu_{j}}(z') A_{m_{j}m}(z,z') \times
\\
\times g_{m_{j}\nu_{j}}(z)\, . 
\label{HFeq}
\end{multline}

\noindent
Here $V_{m}(z)$ represents the effective electron-core potential, $U_{mm'}(z)$ and $A_{mm'}(z)$ denote the effective electron-electron potential and the exchange potential, respectively. Their explicit forms are given in Appendix \ref{app1}.
The energy $E'$ replaces the number of nodes $\nu$ and is characteristic of the unbound electron.

The Hartree-Fock equations \eqref{HFeq} are solved, for every given value
of the continuum energy $E^\prime$, using a standard Runge-Kutta integrator.
Since the $z$ parity is a good quantum number, one
can calculate a symmetric and an antisymmetric wave function by 
imposing corresponding initial conditions at $z=0$. 
For $z\rightarrow\pm\infty$ these wave functions assume a sinusoidal shape, close to the nucleus both the amplitude and the period decrease due to the interaction with the ionic core and the bound electrons.
Because of the delocalised nature of the continuum wave functions 
they can be normalised only over a finite periodicity length $d_z$.
The latter, however, does not enter into the final results

\section{Cross section for the photoionisation}\label{cross}

In this section we give the formula for the photoionisation for the case that both the initial state $\psi_{\text{i}}$ and final state $\psi_{\text{f}}$ are Slater determinants. 
The former is composed of $N$ bound states, the latter consists of $N-1$ bound and one unbound wave function.  The Hamiltonian $\hat{H}_{\text{int}}$ for 
the interaction of the electrons with the radiation field   
$\vec{A}_{\text{r}} ~ (\propto \gvec{\epsilon}\e^{\i\vec{k}\vec{r}})$ of the
incoming photon is obtained by the usual minimal substitution of the momenta,
$ \vec{p}_i \to \gvec{\pi}_i=\vec{p}_i-\sqrt{2}\vec{A}_{\text{r}}$ (recall that in atomic
Rydberg units $e = \sqrt{2}$), and by taking into 
account the interaction of the spin magnetic moments of the electrons 
with the magnetic
field $\vec{B}_{\rm r} = \nabla \times \vec{A}_{\text{r}}$ 
of the radiation field:
\be
\hat{H}_{\text{int}} = \sum_{i=1}^{N}(2\gvec{\beta}_{\text{r}}\vec{\hat{\gvec{\sigma}}}_i 
- 2\sqrt{2}\vec{A}_{\text{r}}{\gvec{\pi}_i})\mathds{1}.
\label{hint}
\ee
\noindent
In \eqref{hint},  $\gvec{\beta}_{\text{r}} = \vec{B}_{\rm r}/B_0$, and since we are 
interested in one-photon transitions the term quadratic in the vector
potential has been omitted. The (one-dimensional) density of final states
of the emitted (spin-down) electron is given by
\be
\rho_{\text{f}}=\frac{d_{z}}{2\pi}\sqrt{\frac{1}{4E'}}\,  \nonumber
\label{zustandsdichte}
\ee
where $d_z$ denotes the periodicity length in $z$-direction.

Starting from Fermi's golden rule for the transition rate $w(\Psi_{\rm f}, \Psi_{\rm i}) = 2 \pi \left | \langle \Psi_{\rm f}|\hat{H}_{\text{int}}|\Psi_{\rm i}\rangle \right |^2 \rho_{\text{f}}$, we can calculate the cross section by the 
usual relation $w = j\sigma$, with $j = c/V$ the photon flux of the incident
radiation. Noting that in atomic Rydberg units $c = 2/\alpha$, 
where $\alpha$ is the fine-structure constant, we have the relation
$\sigma = \frac{V \alpha}{2} w$.

We calculate the photoionisation cross sections in dipole approximation.
Therefore there are  three types of transitions, which differ in the change of the total angular momentum in $z$ direction (quantum number $M=\sum m_i$) caused by the absorption of the photon.
Linearly, right, and left circularly polarised light induces transitions, with $\Delta M=0$ ($\sigma_0$), $\Delta M=+1$ ($\sigma_+$) and  $\Delta M=-1$ ($\sigma_-$), respectively.

It was shown in \citep{Potekhin97} that in the 'length form' the cross sections in adiabatic and non-adiabatic form coincide.
Therefore we also switch from the 'velocity form' to the 'length form' using the well known commutator relation
\be
[\hat{H}_{i},\vec{r}_i]=-2\i\gvec{\pi}_i\, , \nonumber
\label{rpform}
\ee
\noindent
where $\hat{H}_{i}$ designates the non-interaction-Hamiltonian of the $i$-th electron, cf. \eqref{hamiltonzylinder}.

Lengthy but straightforward calculations finally yield explicit expressions for the photoionisation cross sections of the three types of dipole transitions, which can be found in Appendix \ref{app2}.

\section{
Cross sections of two-electron systems
}\label{results}

For reasons of simplicity we choose helium and helium-like oxygen to be the systems of interest to demonstrate our method of calculation. They are still simple enough to compute a huge amount of single cross sections in reasonable time, and are complex enough to draw conclusions for other elements.

\subsection{Single Cross Section}

As an example, we discuss six single cross sections of helium and compare them to previously calculated cross sections \citep{Medin}. We will show that our algorithm can reproduce and improve upon these results. All presented cross sections   
are transitions where the ground state (configuration $(0,0)(1,0)$) is the initial state.
\onecolumn
\begin{figure}[hb]
\centering
\includegraphics[width=0.9\columnwidth]{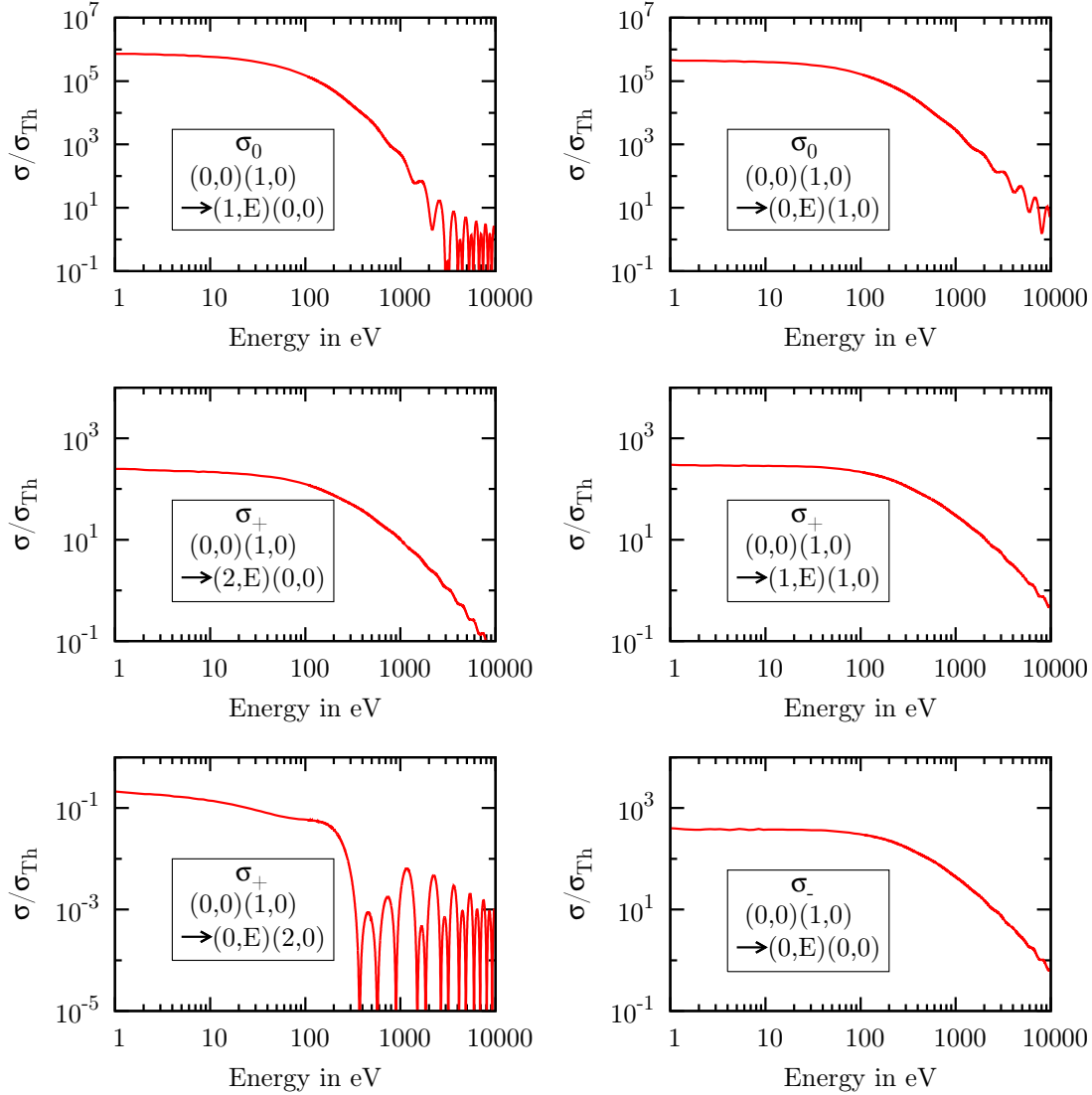}
\caption{(Color online) Single cross sections for the first six transitions from the ground state $(0,0)(1,0)$ with different polarisations. The magnetic filed is $B=10^8$\,T. In comparison to \citep{Medin} these cross sections show more detail for higher energies.}
\label{fig:scs-medin}
\end{figure}

\begin{multicols}{2}
The six cross sections are shown in Figure \ref{fig:scs-medin}, and are very similar to those calculated by Medin (see Figure .. in \citep{Medin}. However, there are differences in the low and high energy regime. For low energies, the values of the cross section are smaller as compared to those calculated by Medin et al.. Furthermore, from a certain energy on, the product of the bound and unbound wave functions becomes increasingly oscillating. This leads to a decline of the cross section and the occurrence of zero crossings. 
\end{multicols}

\twocolumn

We also have to analyse, what terms in the cross sections are important. For now the transition between the initial and the final state is given by $\ket{g^\in_{m_{1} \nu_{1}}} \ket{g^\in_{m_{2} \nu_{2}}}$ and $\ket{g^\f_{m_{3} E'}} \ket{g^\f_{m_{4} \nu_{4}}}$. For two-electron atoms the double sum in \eqref{wqs0} reduces to four possible terms, two with the continuum wave function in the dipole matrix element, and
two with the continuum wave function in the minor:
\be\bs
\brakket{g^\f_{m_{3} E'}}{  z_1 }{g^\in_{m_{1} \nu_{1}}} \delta_{m_{3},m_{1}} a^{\fin}_{31} \\ + \brakket{g^\f_{m_{3} E'}}{  z_2 }{g^\in_{m_{2} \nu_{2}}} \delta_{m_{3},m_{2}} a^{\fin}_{32}\\
+\brakket{g^\f_{m_{4} \nu_{4}}}{ z_1 }{g^\in_{m_{1} \nu_{1}}} \delta_{m_{4},m_{1}} a^{\fin}_{41} \\ +\brakket{g^\f_{m_{4} \nu_{4}}}{ z_2 }{g^\in_{m_{2} \nu_{2}}} \delta_{m_{4},m_{2}} a^{\fin}_{42} 
\end{split}\label{4Terme}
\ee
The behaviour of single cross sections is mainly determined by the unbound wave function. Since $z$-parity is a good quantum number the wave function is either symmetric or antisymmetric.
In the transitions considered both single-particle orbitals of the initial state
are symmetric. Therefore, if the continuum wave function appears in the minor,
as is the case for the last two terms in \eqref{4Terme}, the latter
has a value different from zero only for the symmetric continuum wave function.
On the other hand, does the continuum wave function appear in the
dipole matrix element, as is the case for the first two terms in \eqref{4Terme},
only the antisymmetric continuum wave function yields a nonvanishing
contribution because of the antisymmetry of the operator $z$. So generally only two of the four terms are important for one cross section. 

\begin{figure}[htb]
\includegraphics[width=\columnwidth]{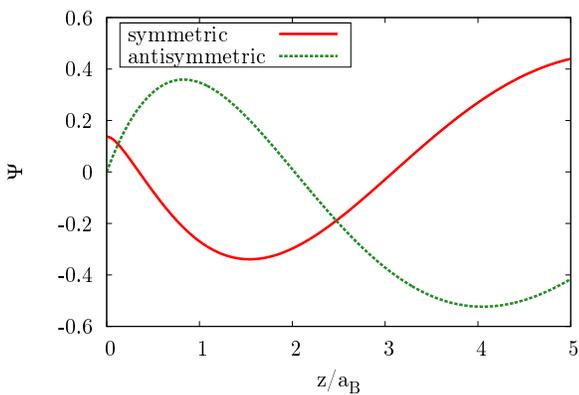}
\caption{\label{DPHef199A0wf} (Color online) Symmetric and antisymmetric continuum wave functions of helium for the state $(m_3=1,E'=0.1~\rm{eV})\ (m_4=0,\nu_4=0)$ at $B=10^8$~T ($\beta\approx 212$).}
\end{figure}

If we take a look at a typical wave function like the one presented in Figure \ref{DPHef199A0wf}, we can see that the form of the symmetric and the antisymmetric continuum wave function near zero is very different. In this region, the value of the cross section is determined by the form of the bound wave function and attains a high value when both the bound and continuum wave function coincide in their shapes.

\subsection{Total Cross Section}

We now wish to calculate the total cross section  for the ionisation by
a photon of a given energy, which is the sum over all single cross sections $\sigma = \sum_i \sigma_i$. Since in the lowest Landau level the magnetic quantum numbers in principle are not bounded below, this sum would diverge. To find a physically reasonable cut-off we make use of the fact that the spatial extension $\rho_0$ of a Landau
state $\Phi_{m}(\rho,\phi)$  is characterized by \citep{CanutoVentura}
\be
\label{extent}
\rho_0^2 = a_L^2 \left ( |m| +\frac{1}{2} \right )
\ee
\ni
with $a_L=a_0/\sqrt{\beta}$ the Larmor radius. This means that as $|m|$ increases the atoms become broader and broader in the direction perpendicular to the field until, at a given mass density, for some $|m|$ they begin to ''touch'' each other. This determines the maximum magnetic quantum number that has to be
considered when summing for the total cross section. In the results presented
below we chose a maximum value of $\abs{m} = 200$. In an analogous fashion
we restrict ourselves to longitudinal wave functions with a maximum
number of nodes $\nu = 2$, since for larger number of nodes the 
orbitals would overlap at a given mass density. In Section~V we will
give values for the mass densities that are covered by these restrictions.

\begin{figure}[htb]
\includegraphics[width=\columnwidth]{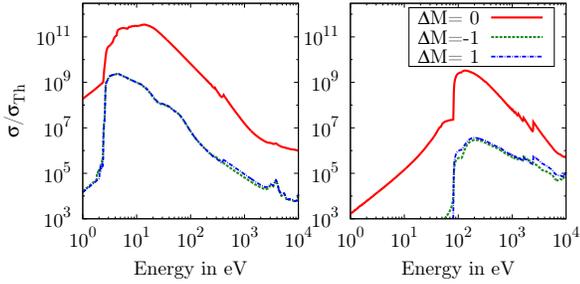}
\caption{(Color online) Left: Total cross section for the photoionisation of helium at $B = 10^8$\,T obtained by summing over all initial 
states with $|m| \le 200$
and $\nu \le 2$ as a function of the photon energy
for the three different polarisations. Right: The corresponding cross section 
for helium-like oxygen.}
\label{fig:tcs-He}
\end{figure}

We note that with this the number of initial states amounts 180,300.
Starting from these states, every possible transition has to be found. For every polarisation, these states lead in total to roughly 1 million transitions. 
In Fig. \ref{fig:tcs-He} we compare the total cross section of helium, summed over all possible ionisations, and helium-like oxygen.

In both cases, the cross sections with circularly polarised light are several orders of magnitude smaller than the cross section with linearly polarised light. Also the general behavior of the curves for both elements is similar. The main difference is the position of the maximum cross section. As can be expected, the maximum for oxygen is at higher energies, due to the higher binding energy. The second difference is the absolute value, which is smaller for oxygen.

\section{Cross section for realistic physical conditions of a neutron star}
\label{influences}

In this section we calculate effective photoionisation cross sections by taking
into account the physical conditions that prevail in the neutron star atmosphere. First we  assume a thermal occupation of the initial states and 
calculate total cross sections as a function of photon energy. 
Then we determine cross sections averaged over the photon energies of
a Planckian spectrum. Finally we explore the effects of the plasma density
and estimate the values for the mass density for which the cut-off
criteria are valid.

\subsection{Thermal occupation probability}
\label{sub:top}
Neutron star atmospheres are hot, with temperatures up to $10^6$~K 
\citep{Zavlin09a}. We account for this fact by a thermal
occupation of the initial states, and in summing the single photoionisation 
cross sections weight their contributions to the total cross section 
according to  their occupation probability $p(E) = e^{-\Delta E/kT}$. 
From \citep{Mori2006a} and \citep{Bignami} we adopt a representative
value of $kT = 150$\,eV and, to demonstrate the influence of an increase
of temperature, of $kT = 300$\,eV. 

\begin{figure}[htb]
 \includegraphics[width=\columnwidth]{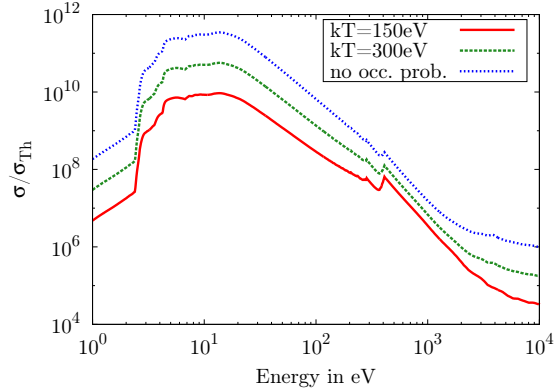}
 \centering
 \caption{(Color online) Total cross section for the photoionisation of helium at $B = 10^8$\,T as a function of the photon energy for linear polarisation obtained by summing over all initial states with $|m| \le 200$
and $\nu \le 2$ and assuming thermal occupation of the initial
states.
The cross section without thermal occupation from Fig.~\ref{fig:tcs-He} is shown for comparison.}
 \label{fig:cross-section-notherm}
\end{figure}

The resulting cross sections are shown in Fig. \ref{fig:cross-section-notherm}.
At higher temperatures the occupation probabilities of the excited bound states are larger, which yields  an increase of the total cross section. Also, without 
thermal occupation weighting the cross section was much larger. We have the effect that single cross sections of excited initial states which manifested themselves at lower energies can be rather large, but are suppressed if thermal occupation probability is taken into account.

\begin{figure}[htb]
 \includegraphics[width=\columnwidth]{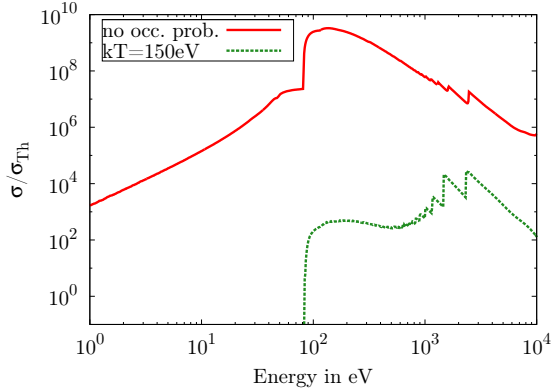}
 \centering
 \caption{(Color online) As Fig.~\ref{fig:cross-section-notherm} but for helium-like oxygen.}
 \label{fig:cross-section-therm-oxygen}
\end{figure}

This effect becomes even more pronounced when we proceed to heavier elements 
such as oxygen (see Fig. \ref{fig:cross-section-therm-oxygen}).  Below 100 eV practically no transitions contribute, and the peak around 100 eV becomes suppressed. The reason is that the energy difference between excited states and the ground state of oxygen is much larger than for helium, therefore the occupation probability is very small. In fact, we can show that there are only 600 transitions left which contribute significantly to the cross section. All other transitions are negligible.

In the following, we concentrate on the total cross section for thermal occupation of the initial states with $kT=150$\,eV. Circularly polarised light is not considered in particular since it shows the same qualitative behavior as linearly polarised light.

\subsection{Photon density}
\label{sub:photondensity}

\begin{figure}[htb]
 \includegraphics[width=\columnwidth]{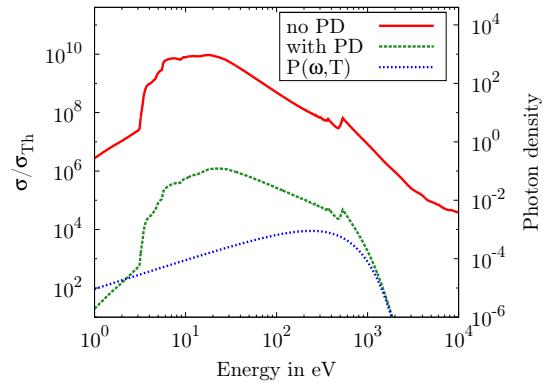}
 \centering
 \caption{(Color online) 
Total cross section for the photoionisation of helium at $B = 10^8$\,T obtained by summing over all initial states with $|m| \le 200$
and $\nu \le 2$ for thermal occupation of the initial
states with $kT = 150$\,eV, and averaged over a blackbody spectrum 
of the incident photon  (linear polarisation)
 with the same temperature (lower dotted curve).
The spectrally not averaged cross section from Fig.~\ref{fig:cross-section-notherm}  (upper curve) is shown for comparison.
}
 \label{fig:norm-pd}
\end{figure}

We now take into account the temperature dependent spectral distribution
of photons in the neutron star atmosphere, which we assume to be blackbody,
and calculate total cross sections averaged over the photon energy distribution. 
This is only a simple model, to illustrate the influence of the energy dependence of the incident photon flux. 
It is these cross sections which are needed for the astrophysical modeling 
of the spectra of thermally emitting magnetised neutron stars.
The starting point is Planck's law divided by the photon frequency and, due to the assumption of a photon flux with one photon per volume element, normalized by the factor
\be
\int_0^\infty \tilde P(\omega,T)\dd \omega = \frac{4 \pi T^3}{\alpha^3} 2 \zeta(3) \ ,
\ee
\ni
where $\zeta(x)$ is the Riemann zeta function. This yields the photon density
\be
P(\omega,T) \dd \omega = \frac{\omega^2}{2\ZETA(3) T^3} \frac{1}{\e^{\omega/ T} -1}  \dd \omega \, .
\label{photondensity}
\ee
\ni
The resulting photon density is shown in Fig. \ref{fig:norm-pd}, for the 
temperature of $kT=150$\,eV.

The effect of the photon density is best shown in a comparison between the cross sections without photon density and the same cross section multiplied by the photon density $P(\omega)$ (Fig. \ref{fig:norm-pd}). The cross section changes its qualitative behavior towards the shape of the photon density. The peaks at energies around 20\,eV are suppressed and the decay at energies greater than 400\,eV is more rapid due to the low photon density. 

\begin{figure}[htb]
 \includegraphics[width=\columnwidth]{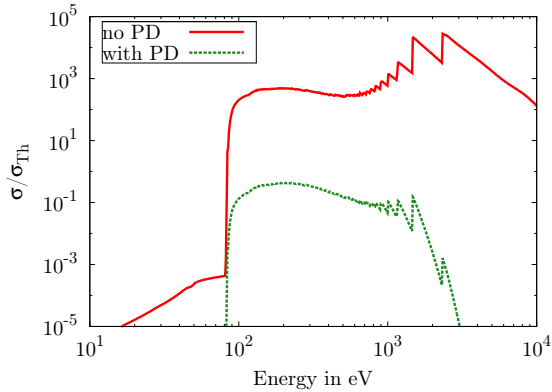}
 \centering
 \caption{(Color online) As Fig.~\ref{fig:norm-pd} but for helium-like oxygen
at $B = 10^8$\,T.}
 \label{fig:pd-oxygen}
\end{figure}

The same picture for helium-like oxygen shows intriguing behavior (Fig. \ref{fig:pd-oxygen}). The last two peaks around 2 keV originally have about the same value. When the photon density is taken into account the last peak is two orders of magnitude smaller. Thus the photon density can change the the shape of 
the final total cross section significantly.

\subsection{Density of the plasma}
\label{sub:plasmadensity}

The atmosphere of a neutron star consists of ionised atoms. The volume an
atom can fill is restricted by the particle density.
Because of the enormous gravity at the surface neutron star 
atmospheres are strongly suppressed, and values for the mass density
one finds in the literature range between  
 $\varrho_p = 1$\,g/cm$^3$ and $\varrho_p = 100$\,g/cm$^3$ (see e.g. \citep{Mori2006a,Zavlin09a}).
Assuming that all atoms occupy the same state, the average volume 
available per atom $V_s$ can be written as
$V_s = {m_{\text{atom}}}/{\varrho_p}$
where $m_{\text{atom}}$ is the atomic mass. No overlap of adjacent atoms with volumes smaller than $V_s$ exists, and we have included only the contributions
from these atomic states. We now wish to estimate for which mass densities
in our calculations the restriction  to $\abs{m} \le 200$ and $\nu \le 2$
is justified, and whether or not for higher mass densities the maximum
$\abs{m}$ can even be decreased.

We make the simple approximation that the atoms are cylindrically shaped 
and aligned gaplessly. As already noted the spatial extension $\rho_0$ of a state perpendicular to the magnetic field is given by  \eqref{extent}.
We only consider the electron orbital with the largest extent, that is, with
the maximum $\abs{m}$. The corresponding volume of the atom is
$
V_a =  2\pi \rho^2_0(|m|) \sqrt{\ew{z^2}_{\text{max}}} \ ,
$
where $\sqrt{\ew{z^2}_{\text{max}}}$ denotes the maximum extension in $z$-direction. Generally, higher $\nu$ result in larger extensions. If the calculated volume $V_a$ is bigger than $V_s$, the state is omitted as an initial state. The same is true for final states. 

In a more detailed treatment one would have to consider that the atoms are 
not purely cylindrical but cigar-shaped and therefore, more or less, aligned in an ellipsoidal packing. Moreover different states coexist, and the atoms are not aligned gaplessly but exhibit a separation. However, our simple approximation is good enough for the estimates that we are seeking For a more realistic approach, see e.g. \citep{mihalas}.

\begin{table}[htb]
 \begin{tabular}{c  c  c  c  }
  \toprule
  \multicolumn{1}{c}{$\varrho$ in $\text{g}/\text{cm}^3$} & \multicolumn{1}{c}{$\nu$} & \multicolumn{1}{c}{highest $\abs{m}$} & \multicolumn{1}{c}{$N_t$} \\
  \midrule
    & 0 & 199 & \\ 
  1 & 1 & 169 & 336\,172 \\ 
    & 2 & 125 & \\ 
    \cmidrule(l{5pt}r{5pt}){1-4}
    & 0 & 36 & \\ 
  10& 1 & 22 &  4\,978 \\ 
    & 2 & 15 & \\ 
    \cmidrule(l{5pt}r{5pt}){1-4}
    & 0 & 20 & \\ 
  20& 1 & 11 & 1\,400 \\ 
    & 2 & 7 & \\ 
    \cmidrule(l{5pt}r{5pt}){1-4}
    & 0 & 9 & \\ 
  50& 1 & 4 & 282 \\ 
    & 2 & 2 & \\ 
    \cmidrule(l{5pt}r{5pt}){1-4}
    & 0 & 5 & \\ 
 100& 1 & 2 & 71 \\ 
    & 2 & 1 & \\ 
    \bottomrule
 \end{tabular}
 \centering
 \caption{Highest $\abs{m}$ depending on $\nu$ at different plasma densities and the total number of transitions $N_t$.}
 \label{tab:maxm}
\end{table}

In Table \ref{tab:maxm} we list, for five different mass densities the maximum $\abs{m}$ depending on $\nu$ for which the volume of this state is still small enough to contribute to the cross section. This maximum quantum number declines fast for higher densities. It is interesting to note that some volumes of states with $\nu=2$ at higher densities are still small enough. This is due to the very small $\rho_0$ if $m=0$, and for two electrons the interaction with the nucleus is very strong so that $\ew{z^2}$ increases only slowly. Therefore states with $m=0$ can still be small enough, even if $\nu$ becomes very large. This also 
implies, that we still do not have a complete total cross section, but due to the low $\abs{m}$ and $\nu$ seen in Table \ref{tab:maxm}, it is possible to calculate all states and transitions in reasonable time.

 The Table also  
shows that the number of transitions that have to be considered 
when calculating the total cross section. This number
decreases rapidly, until at 100\,g/cm$^3$ there is only 0.006\% of the transitions necessary originally left. 

\begin{figure}[htb]
 \includegraphics[width=\columnwidth]{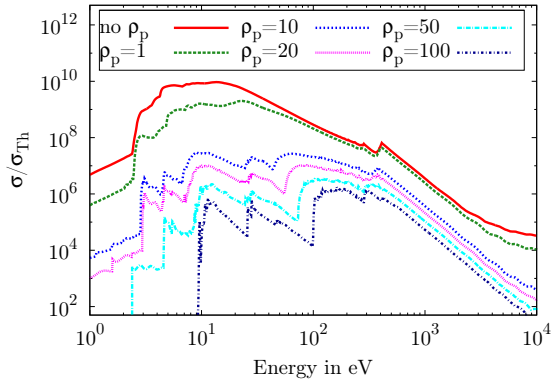}
 \centering
 \caption{(Color online) Total photoionisation cross section of helium
at $B = 10^8$\,T for linear polarisation obtained by summing only over
as many initial states as are compatible with the mass density $\varrho_p$
(cf. Table~\ref{tab:maxm}). Values of $\varrho_p$  are given in 
$[\text{g}/\text{cm}^3]$. Thermal occupation of the initial states 
with $kT=150$\,eV is assumed. The 
thermally averaged cross section from Fig.~\ref{fig:cross-section-notherm}
in which the mass density is not taken into account 
(top most curve) is shown for comparison. It can be seen 
 that as the density increases the cross sections significantly change in height and shape.}
 \label{fig:plasmadensity}
\end{figure}

Figure \ref{fig:plasmadensity} shows the cross section, calculated for linearly polarised light, with mass densities between 1 and 100\,g/cm$^3$.
It can be seen that at 1\,g/cm$^3$ the cross section changes only quantitatively, but not its qualitative behavior, while the number of transitions decreases by 740,000, which is 70\% of the total transitions. The consequence is that many single cross section are negligible.

Starting at 10 g/cm$^3$, the cross section is significantly changed. There is only a fraction of the originally calculated cross sections left, which is the reason for this huge change.

In particular, mostly highly excited states drop out. These transitions contribute to the lower energy range (1-50\,eV) and this is why the cross section changes significantly in this range.

\subsection{
Finite Core Mass
}
\label{sub:fcm}

In all our calculations we have worked in the infinite-nuclear-mass approximation. We therefore 
finally discuss the influence of the finite nucleus mass on the results.
It has been shown \citep{Wunner1980,Ruder} that the single-particle levels are raised by 
the cyclotron energy $\hbar \omega^+$ of the charged atomic nucleus. Intuitively
this corresponds to  its own gyration as a charged particle in the magnetic field. 

This so called $m$-shift rises the energy levels of every state, so that many of these are shifted over the ionisation edge, and therefore need no longer be accounted for in the total cross section.

For the magnetic field strength considered here ($B = 10^8$~T), $\hbar \omega^+ = 1.67$~eV for helium,
and $0.39$~eV and $0.11$~eV for helium-like oxygen and iron, respectively.
This does not influence transition energies with $\Delta M$ = 0, 
since all levels are raised by the same amount, and gives
only a small correction for the, as compared to $\Delta M = 0$, weaker
transitions with $\Delta M = \pm 1$.

\begin{figure}[htb]
 \includegraphics[width=\columnwidth]{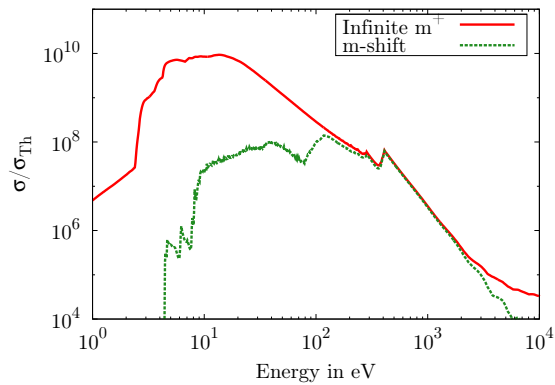}
 \centering
 \caption{(Color online) The cross section for helium at $B=10^8$\,T for linear polarisation and thermal weight at $kT=150$\,eV with account for the finite nuclear mass. For comparison the total cross section with infinite nuclear mas is shown (upper curve).}
 \label{fig:finitemass}
\end{figure}

Figure \ref{fig:finitemass} shows the total cross section calculated with finite nuclear mass. As was expected, the highly excited states are no longer accounted for, so mainly the low level transitions are dropping out. This behaviour is similar to the mass density effect.

The reduction of the number and composition of transitions is important when calculating opacities. If oversized and autoionising atoms are neglected, we know that no transitions are missing in the total cross section. 

\begin{figure}[htb]
 \includegraphics[width=\columnwidth]{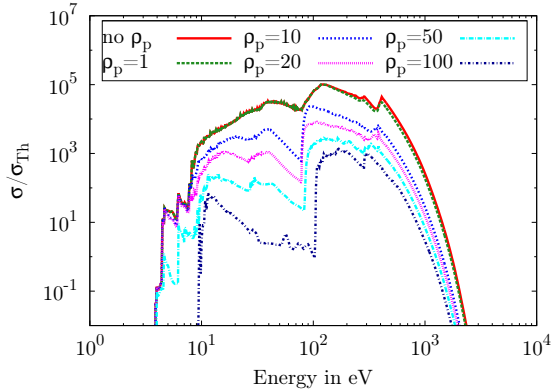}
 \centering
 \caption{(Color online) As Fig.~\ref{fig:plasmadensity} but additionally
with spectral averaging over a blackbody spectrum with $kT = 150$\,eV.
Values of $\varrho_p$ are again given in $[\text{g}/\text{cm}^3]$.}
 \label{fig:plasmadensitypdcorr}
\end{figure}

All physical conditions prevailing in the neutron star atmosphere are taken into account in Fig. \ref{fig:plasmadensitypdcorr}, for a thermal occupation probability at $kT = 150$\,eV, the photon density $P(\omega)$ with $kT = 150$\,eV, different plasma densities and finite nuclear mass. The Figure shows that the peaks at energies between 10 and 50\,eV are suppressed due to the low photon density, and the overall shift towards the shape of the photon density is obvious.

We can again look at the number of transitions remaining for different plasma densities in Table \ref{tab:not}. Here we are only interested in the total number, without the splitting into $\nu$-channels. A comparison quickly shows, that only few atoms that are not oversized drop out of the calculation. This does not mean that finite nuclear mass is negligible, but both effects have to be taken into account, for a complete cross section.

\begin{table}[htb]
 \begin{tabular}{c  c  }
  \toprule
  \multicolumn{1}{c}{$\varrho$ in $\text{g}/\text{cm}^3$} & \multicolumn{1}{c}{$N_t$} \\
  \midrule
  1 & 44967 \\
  10 & 4506 \\
  20 & 1264 \\
  50 & 251 \\
  100 & 59 \\
    \bottomrule
 \end{tabular}
 \centering
 \caption{The total number of transitions $N_t$ at different plasma densities with finite nuclear mass.}
 \label{tab:not}
\end{table}

\section{Summary}

We have developed a program code which allows  calculating
bound-free  transitions of few-electron atoms and ions in neutron star
magnetic fields in a routine way. For the examples
of helium and helium-like oxygen we have analysed, for realistic physical 
parameters,
which transitions contribute significantly to the total photoionisation
cross sections and which are negligible. In this way we have succeeded 
in drastically reducing the numbers of transitions that have to be considered.
The strategies developed in reducing the number of contributing transitions
can also be applied in the calculation of the photoionisation
of atoms and ions with more than two electrons. 

The present results clearly demonstrate the complexity of ab-initio
calculations of realistic photoionisation cross sections in neutron star
magnetic fields 
that can be used in 
astrophysical modeling. An alternative would be the development
of phenomenological models for the cross sections. Here our results
could serve as useful starting point for developing such models.

\section{Acknowledgements}

We  thank the bwGRiD project
(\url{http://www.bw-grid.de}), member of the {G}erman {D}-{G}rid initiative,
      funded by the {M}inistry for {E}ducation and {R}esearch ({B}undesministerium f\"ur
      {B}ildung und {F}orschung) and the {M}inistry for {S}cience, {R}esearch and {A}rts
      {B}aden-{W}\"urttemberg ({M}inisterium f\"ur {W}issenschaft, {F}orschung und {K}unst
      {B}aden-{W}\"urttemberg)
for providing 
the computational resources, 
and Christoph Schimeczek and Sebastian Boblest for valuable discussions 
and input.

\appendix

\section{Potentials}
\label{app1}

The explicit forms of the potetials in \eqref{HFeq} read

\begin{align}
 V_{m}(z)= & -2Z\int \d\vec{r}_{\bot}\frac{|\Phi_{m}(\vec{r}_{\bot})|^{2}}{|\vec{r}|} \, , \nonumber \\
 U_{mm'}(z,z')= & \ 2\iint \d\vec{r}_{\bot} \d\vec{r}_{\bot}^{'} \frac{|\Phi_{m}(\vec{r}_{\bot})|^{2}|\Phi_{m'}(\vec{r}_{\bot}^{'})|^{2}}{|\vec{r}^{'}-\vec{r}|}\, , \nonumber \\ 
 A_{mm'}(z,z')= & \ 2\iint \d\vec{r}_{\bot} \d\vec{r}_{\bot}^{'} \frac{\Phi_{m'}^{*}(\vec{r}_{\bot})\Phi_{m}(\vec{r}_{\bot})\Phi_{m}^{*}(\vec{r}_{\bot}^{'})\Phi_{m'}(\vec{r}_{\bot}^{'})}{|\vec{r}^{'}-\vec{r}|}\, . \nonumber
\end{align}

These three potentials can be computed numerically by evaluating the integrals. For a detailed description of the evaluation of the above integrals we refer to \citep{Proeschel}.

\section{Cross Section}
\label{app2}

The photoionization cross section of the three types of dipole transitions are

\begin{align}
\begin{split}
\sigma_0 =& \frac{3}{8} \frac{d_z \sigma_\text{Th} \omega_{\fin}^2}{\alpha^3 \omega_{\text{ph}}} \sqrt{\frac{1}{E'}}\,  \epsilon_{z}^2
\left | \, \sum_{k=1}^{N-1} \sum_{l=1}^{N} \brakket{g^\f_{m_{k} \nu_{k}}}{  z_l }{g^\in_{m_{l} \nu_{l}}} \delta_{m_{k},m_{l}} a^{\fin}_{kl}  \right.
\\
& \phantom{\frac{3}{8} \frac{d_z \sigma_\text{Th} \omega_{\fin}^2}{\alpha^3 \omega_{\text{ph}}} \sqrt{\frac{1}{E'}}\,  \epsilon_{z}^2 \left | \, \sum_{k=1}^{N-1} \right.} + \left. \sum_{l=1}^{N} \brakket{g^\f_{m_{N} E'}}{ z_l }{g^\in_{m_{l} \nu_{l}}} \delta_{m_{N},m_{l}} a^{\fin}_{Nl}\, \right |^2
\\
\shoveleft
\sigma_- =& \frac{3}{8} \frac{d_z \sigma_\text{Th}\omega_{\fin}^2}{\alpha^3 \omega_{\text{ph}}} \sqrt{\frac{1}{E'}} \abs{\epsilon_+}^2 
\left |\, \sum_{k=1}^{N-1} \sum_{l=1}^{N} \brakket{g^\f_{m_{k}{ \nu_{k}}}}{  \sqrt{\frac{-m_l+1}{4 \beta}} }{g^\in_{m_{l} \nu_{l}}} \delta_{m_{k}-1,m_{l}} a^{\fin}_{kl} \right. 
\\
& \phantom{\frac{3}{8} \frac{d_z \sigma_\text{Th} \omega_{\fin}^2}{\alpha^3 \omega_{\text{ph}}} \sqrt{\frac{1}{E'}}\,  \epsilon_{z}^2 \left | \, \sum_{k=1}^{N-1} \right.} + \left. \sum_{l=1}^{N} \brakket{g^\f_{m_{N} E'}}{  \sqrt{\frac{-m_l+1}{4 \beta}} }{g^\in_{m_{l} \nu_{l}}} \delta_{m_{N}-1,m_{l}} a^{\fin}_{Nl}\, \right |^2
\\
\sigma_+ =& \frac{3}{8} \frac{d_z \sigma_\text{Th}\omega_{\fin}^2}{\alpha^3 \omega_{\text{ph}}} \sqrt{\frac{1}{E'}} \abs{\epsilon_-}^2 
\left |\, \sum_{k=1}^{N-1} \sum_{l=1}^{N} \brakket{g^\f_{m_{k} \nu_{k}}}{ \sqrt{\frac{-m_l}{4\beta}} }{g^\in_{m_{l} \nu_{l}}} \delta_{m_{k}+1,m_{l}} a^{\fin}_{kl} \right.
\\
& \phantom{\frac{3}{8} \frac{d_z \sigma_\text{Th} \omega_{\fin}^2}{\alpha^3 \omega_{\text{ph}}} \sqrt{\frac{1}{E'}}\,  \epsilon_{z}^2 \left | \, \sum_{k=1}^{N-1} \right.} + \left. \sum_{l=1}^{N} \brakket{g^\f_{m_{N} E'}}{  \sqrt{\frac{-m_l}{4\beta}} }{g^\in_{m_{l} \nu_{l}}} \delta_{m_{N}+1,m_{l}} a^{\fin}_{Nl}\, \right |^2   \,. 
\label{wqs0} 
\end{split}
\end{align}

In \eqref{wqs0} $\omega_{\text{ph}}$ is the energy of the incident photon and $\sigma_\text{Th}=8\pi \alpha^4/3$ is the Thomson cross section. 

The subdeterminant $a_{kl}^{\fin}$ is defined by $a_{kl}^{\fin} = (-1)^{k+l} \det(A_{kl})$, where $\det (A_{kl})$ is the minor (remove row $k$ and column $l$) of the matrix consisting of the elements $S_{kl}^{\fin} = \braket{\psi_{\f,k}}{\psi_{\in,l}}$, which represent the overlaps between the single-particle-orbitals  $\psi_{\in,l}$ and  $\psi_{\f,k}$.

A few remarks are in order regarding equation \eqref{wqs0}.
Since we do not take multi-photon transitions into account, we can replace the photon's energy $\omega_{\text{ph}}$ by the energy difference between final and initial state $\omega_\fin$ ($\omega_{\text{ph}}=\omega_\fin$ for single-photon transitions).
Although the periodicity length $d_z$ occurs in the formulae, the calculated cross sections are independent of $d_z$. This is due to the fact that the
continuum wave function of the ionised electron is normalised with respect to
$d_z$ and therefore contains a factor $1/\sqrt{d_z}$.

\bibliographystyle{apj}
\bibliography{literatur}

\end{document}